\documentclass[12pt,epsf]{article}
\usepackage{graphicx}
\begin{document}
%%%%%%%%%%%%%%%%%%%%%%%%%%%%%%%%%%%%%%%%%%%

\def\a{\alpha}
\def\b{\beta}
\def\c{\varepsilon}
\def\d{\delta}
\def\e{\epsilon}
\def\f{\phi}
\def\g{\gamma}
\def\h{\theta}
\def\k{\kappa}
\def\l{\lambda}
\def\m{\mu}
\def\n{\nu}
\def\p{\psi}
\def\q{\partial}
\def\r{\rho}
\def\s{\sigma}
\def\t{\tau}
\def\u{\upsilon}
\def\v{\varphi}
\def\w{\omega}
\def\x{\xi}
\def\y{\eta}
\def\z{\zeta}
\def\D{\Delta}
\def\G{\Gamma}
\def\H{\Theta}
\def\L{\Lambda}
\def\F{\Phi}
\def\P{\Psi}
\def\S{\Sigma}

\def\o{\over}
\def\beq{\begin{eqnarray}}
\def\eeq{\end{eqnarray}}
\newcommand{\gsim}{ \mathop{}_{\textstyle \sim}^{\textstyle >} }
\newcommand{\lsim}{ \mathop{}_{\textstyle \sim}^{\textstyle <} }
\newcommand{\vev}[1]{ \left\langle {#1} \right\rangle }
\newcommand{\bra}[1]{ \langle {#1} | }
\newcommand{\ket}[1]{ | {#1} \rangle }
\newcommand{\EV}{ {\rm eV} }
\newcommand{\KEV}{ {\rm keV} }
\newcommand{\MEV}{ {\rm MeV} }
\newcommand{\GEV}{ {\rm GeV} }
\newcommand{\TEV}{ {\rm TeV} }
\def\diag{\mathop{\rm diag}\nolimits}
\def\Spin{\mathop{\rm Spin}}
\def\SO{\mathop{\rm SO}}
\def\O{\mathop{\rm O}}
\def\SU{\mathop{\rm SU}}
\def\U{\mathop{\rm U}}
\def\Sp{\mathop{\rm Sp}}
\def\SL{\mathop{\rm SL}}
\def\tr{\mathop{\rm tr}}

\def\IJMP{Int.~J.~Mod.~Phys. }
\def\MPL{Mod.~Phys.~Lett. }
\def\NP{Nucl.~Phys. }
\def\PL{Phys.~Lett. }
\def\PR{Phys.~Rev. }
\def\PRL{Phys.~Rev.~Lett. }
\def\PTP{Prog.~Theor.~Phys. }
\def\ZP{Z.~Phys. }

%%%%%%%%%%%%%%%%%%%%%%%%%%%%%%%%%%%%%%%%%%%%%%%%%%%%%%%%%%%%%%%%%%%%

\baselineskip 0.7cm

\begin{titlepage}

\begin{flushright}
IPMU 08-0021\\
KEK-TH-1243
\end{flushright}

\vskip 1.35cm
\begin{center}
{\large \bf
 Upperbound on Squark Masses in Gauge-Mediation Model\\
with Light Gravitino   
}
\vskip 1.2cm
Junji Hisano$^{(a,b)}$, Minoru Nagai$^{(c)}$,  Shohei Sugiyama$^{(a)}$, and T. T. Yanagida$^{(d,b)}$
\vskip 0.4cm

{$(a)$ \em Institute for Cosmic Ray Research (ICRR), \\
University of Tokyo, Kashiwa, Chiba 277-8582, Japan}\\
{$(b)$ \em Institute for the Physics and Mathematics of the Universe (IPMU), \\
University of Tokyo, Kashiwa, Chiba 277-8568, Japan}\\
{$(c)$ \em Theory Group, KEK, Oho 1-1, Tsukuba, Ibaraki 305-0801, Japan }\\
{$(d)$ \em  Department of Physics, University of Tokyo,\\
     Tokyo 113-0033, Japan}

\vskip 1.5cm

\abstract{
We examine stabilities of our supersymmetry-breaking false vacuum in a low-energy direct 
gauge mediation model of SUSY breaking. The stability required in the high-temperature early universe
leads to upperbounds on  masses of squarks and gluino as $m_{\tilde{q}} \lsim 1~{\rm TeV}$
and $m_{\tilde{g}} \lsim 1~{\rm TeV}$ for the light gravitino of mass $m_{3/2}\lsim 16$~eV. 

 }
\end{center}
\end{titlepage}

\setcounter{page}{2}

\section{Introduction}

The light gravitino of mass $\lsim 16$~eV is very attractive, since
it causes no astrophysical and cosmological
 \cite{Viel:2005qj,Kawasaki:2004yh,Kawasaki:2006gs} gravitino problem.  The gauge mediation
\cite{Dine:1981za} is only a known mechanism to have such
a light gravitino with a consistent spectrum for supersymmetry (SUSY)
particles.  The light gravitino requires relatively light messenger
quarks and leptons at $O(100)$~TeV for generating SUSY particles of
mass $O(1)$~TeV. This is an attractive point for future high-energy
experiments. In addition, the latest experimental data on the
anomalous magnetic moment of the muon and on the branching ratio of
$\bar{B}\to X_s\gamma$ favor a low messenger scale ($\sim 100$~TeV)
\cite{Hisano:2007ah}. However, on the other hand, 
the presence of the light messengers may cause a serious problem
in cosmology, since it often
generates a SUSY-invariant true vacuum near the SUSY-breaking false
one.

If there is indeed a SUSY-invariant true vacuum near our SUSY-breaking
false vacuum, the vacuum transition to the true one should be
sufficiently suppressed at least in the present universe.  In a
recent article \cite{Hisano:2007gb}, constraints from the quantum
vacuum transition have been derived in a generic minimal gauge-mediation
model. However, one should also consider the thermal vacuum
transition if the universe underwent through a high temperature
regime.  The purpose of this paper is to discuss the vacuum transition
at high temperatures in the early universe by assuming an explicit model
for the gauge mediation. For this purpose we take the IYIT
SUSY-breaking model \cite{Izawa:1996pk}, since it easily accommodates
a gauge-mediation model with the light gravitino.

In this paper we evaluate both the quantum and thermal transition
rates of the SUSY-breaking to SUSY-invariant true vacua in the
model. We derive upperbounds on squark and gluino masses
($m_{\tilde{q}} \lsim 1~{\rm TeV}$ and $m_{\tilde{g}} \lsim 1~{\rm
  TeV}$) for the stability of our SUSY-breaking vacuum in the early
universe. The bounds are given in the minimal messenger model. We also
briefly discuss the upperbounds on masses for the SUSY particles in
possible extensions of the minimal model.

\section{Gauge-mediation model with light gravitino}

Our model is based on the IYIT model \cite{Izawa:1996pk} for SUSY
breaking. We introduce a hidden $SU(2)$ gauge group with four chiral
superfields $Q_i$ ($i=1\dots4$) transforming as fundamental
representations of $SU(2)$. We add singlets $S^{ij} (=-S^{ji})$ in
${\bf 6}$-dimensional representation of the flavor $SU(4)$ symmetry of
our model. This particle content is minimal in the IYIT SUSY-breaking
models. The superpotential is given by $W = - \frac12 \lambda
S^{ij}Q_iQ_j$.  At the low energy this theory is in the confining
phase and the moduli space is modified by the strong dynamics as,
${\rm Pf} (Q_iQ_j)=\Lambda^4$, where $\Lambda$ denotes the holomorphic
dynamical scale.

The low-energy effective superpotential is dictated with singlets
$S^{ij}$ and mesons $M_{ij} \simeq Q_i Q_j$. It is convenient to
exploit the local equivalence of $SU(4)$ and $SO(6)$ for the flavor
symmetry, and to regard both singlets and mesons to be in the vector
representations of $SO(6)$, $S_a$ and $M_a$ $(a=0\cdots 5)$. In those
representations, the quantum modified moduli
space and superpotential are given as 
\begin{equation}
{\rm Pf}(M)=M_aM_a =\Lambda^2 \ ,
\end{equation}
\begin{equation}
W_{\rm eff} =- \lambda  \Lambda S_a M_a \ ,
\end{equation}
respectively.  Here we have rescaled by factor of $\Lambda$ so that
$M_a$ has mass dimension $+1$.  By solving the constraint of the
quantum moduli space, we get
\begin{eqnarray}
W_{\rm eff} &=&
-\lambda \Lambda S \sqrt{\Lambda^2-M_a^{\prime}M_a^{\prime}}-
\lambda \Lambda S_a^{\prime}M_a^{\prime} \ ,
\end{eqnarray}
where $S=S_0$ and $M_a^\prime =M_a$ and $S_a^\prime =S_a$ for $a=1\cdots 5$.
SUSY is dynamically broken with 
\begin{equation}
  F_S (\equiv \mu^2) = \lambda \Lambda^2 \ .
\end{equation}

The most important ingredient for our purpose is the form of Kahler
potential. The integration of the light mesons $M_a^{\prime}$ and singlets
$S_a^\prime$ gives rise
to the effective Kahler potential,
\begin{eqnarray}
  K = S^\dagger S - \frac{\eta}{4\Lambda^2} (S^\dagger S)^2 + \cdots \ ,
\label{kahler}
\end{eqnarray}
around $S\lsim 4\pi \Lambda$.  The parameter $\eta$ is \cite{Chacko:1998si}
\begin{eqnarray}
\eta \simeq \frac{5 \lambda^2}{16\pi^2}(2\log2-1)
\end{eqnarray}
at the one-loop level.  The second term in Eq.~(\ref{kahler}) gives a mass
term to the scalar component of $S$ as $m_S^2=
\eta|F_s|^2/\Lambda^2=\eta \lambda \mu^2$, and it makes the
SUSY-breaking vacuum classically stable around $S\simeq 0$ since $\eta>0$. 
Other hadron states also contribute to the Kahler potential, which is, however, 
uncalculable due to the strong dynamics of the hidden $SU(2)$. Fortunately,
it is found,
using the naive dimensional analysis, that the contribution from the light
mesons and singlets to $m_S^2$ in Eq.~(\ref{kahler}) dominates over those from the
uncalculable hadron integrations \cite{Chacko:1998si}.

We introduce a minimal messenger sector. It contains only a pair of
messenger quarks and leptons, $\Phi_{ d}+{\bar \Phi}_{{d}}$ and
$\Phi_{l}+{\bar \Phi}_{l}$, where they belong to $ {\bf 5}+{\bf
  5^\star}$ in the $SU(5)_{\rm GUT}$, together. We do not introduce
more messengers, since it may cause too large CP violation.  In this
sense, an alternative choice of the messengers is a pair of messengers
belonging to ${\bf 10}+{\bf 10^\star}$.  We will shortly discuss
effects of this choice later.

The superpotential for the messengers $\Phi_i$ and ${\bar \Phi_i}$ $(i=d,~l)$ is
\begin{eqnarray}
W= (\kappa_d S-M_d)\Phi_{d} {\bar \Phi}_{d}+ (\kappa_l S-M_l)\Phi_{l}{\bar \Phi}_{l} \ .
\label{messenger}
\end{eqnarray}
In this paper we take $M_d=M_l(\equiv M)$ and $\kappa_d=\kappa_l(\equiv
\kappa)$ for simplicity.  One-loop diagrams generate gaugino masses in
the SUSY standard model (SSM), $ M_{\tilde{g}_i}$ $(i=1$-$3)$, as
\begin{eqnarray}
 M_{\tilde{g}_i}
  &=& \frac{\alpha_i}{4\pi}\frac{\kappa \mu^2}{M} ~g(x),
\end{eqnarray}
and the two-loop diagrams give sfermion masses $m_{\tilde{f}}$ as
\begin{eqnarray}
m_{\tilde{f}}^2&=&
2\left[
C_3^{\tilde{f}} \left(\frac{\alpha_3}{4\pi}\right)^2 
+C_2^{\tilde{f}} \left(\frac{\alpha_2}{4\pi}\right)^2 
+\frac35 Y_{\tilde{f}}^2 \left(\frac{\alpha_1}{4\pi}\right)^2 
\right] 
\left(\frac{\kappa \mu^2}{M}\right)^2~f(x)
\ ,
\label{eq:squark_mass}
\end{eqnarray}
where $C_{3}^{\tilde{f}}$, $C_{2}^{\tilde{f}}$ and $Y_{\tilde{f}}$ are
$SU(3)_C$ and $SU(2)_L$ quadratic Casimirs and hypercharge of sfermion
$\tilde{f}$, respectively. The mass functions $f(x)$ and $g(x)$ depend
on $x\equiv \kappa \mu^2/M^2$, and their explicit forms are given in
Ref.~\cite{Martin:1996zb}.

Now we discuss classical stability of the SUSY-breaking vacuum in our model.
Introduction of the messengers generates the SUSY-invariant true vacuum at
\begin{eqnarray}
S\simeq\frac{M}{\kappa},~~~\Phi_{d}{\bar \Phi}_{d}+\Phi_{l}{\bar \Phi}_{l}=
\frac{\mu^2}{\kappa} \ .
\end{eqnarray}
The SUSY-breaking false vacuum is, however, classically stable under following two
conditions. First, masses for the bosonic components of the
messengers are given by $M^2\pm \kappa \mu^2$ around
$S\simeq 0$. Then, the SUSY-breaking vacuum is classically stable when
\begin{eqnarray}
M^2\gsim \kappa \mu^2 \ .
\label{cond1}
\end{eqnarray}
Second, the mass terms of the messengers in Eq.~(\ref{messenger})
break the $U(1)_R$ symmetry, and it generates corrections to the
scalar potential for $S$, which may make the SUSY-breaking vacuum
unstable \cite{Murayama:2007fe}. One-loop diagrams with the messengers
give the corrections to the scalar potential for $S$ as
\begin{eqnarray}
\delta V\simeq 
-\frac{5 \kappa^2 \mu^4}{16\pi^2}
\left[\frac{\kappa}{M}(S+S^\dagger)
+\frac{\kappa^2}{2 M^2}(S^2+S^{\dagger2})
+\cdots\right] \ .
\end{eqnarray}
The SUSY-breaking minimum is sifted toward the SUSY-invariant vacuum $(\sim
(\kappa/\lambda)^3\times(\mu^2/M))$, and the false vacuum becomes
destabilized unless
\begin{eqnarray}
\lambda^3\gsim \kappa^{4} \frac{\mu^2}{M^2} \ .
\label{cond2}
\end{eqnarray}

We see that the classical stability conditions of the SUSY-breaking false vacuum
give a lowerbound on the gravitino mass. The gravitino mass is given
by
\begin{eqnarray} 
 m_{3/2} \simeq \frac{F_S}{\sqrt{3}M_{\rm pl}}
 \simeq 10~{\rm eV} \left(\frac{m_{\tilde q}}{2~\rm TeV}\right)^2
                   \left(\frac{M}{\kappa \mu}\right)^2\  ,
\label{eq:gravitino}
\end{eqnarray}
where $M_{\rm pl} = 2.4 \times 10^{18} $~GeV is the reduced Planck
mass.  Here the mass scale is normalized by the squark mass $m_{\tilde
  q}$, which is approximately given by Eq.~(\ref{eq:squark_mass}) with
$f(x)=1$. This is because the mass function $\sqrt{f(x)}$ is less
sensitive to $x$ than $g(x)$
($g(x)$ changes gradually from $1$ to $1.4$). It is found
from Eqs.~(\ref{cond1}, \ref{cond2}) that the last factor in
Eq.~(\ref{eq:gravitino}) is larger than $\sim \kappa^{-1}$ and also
$\sim {\kappa}^2/{\lambda}^3$.  Larger $\lambda$ is favored for a
smaller gravitino mass.  On the other hand, the reliability of our
perturbative calculation requires $\lambda \lsim 2$.  Notice that if one imposes
that the coupling
$\lambda$ is small enough for perturbative calculation to be valid
at the GUT scale and no extra $SU(2)$ multiplets are introduced,
$\lambda$ should be smaller than $\sim 1$ at the SUSY-breaking scale. Thus, 
the light gravitino mass
$m_{3/2}<16~{\rm eV}$ is already marginal even from a viewpoint of the
classical stability conditions in our model.

In the following, we evaluate the quantum and thermal transition
rates from the SUSY-breaking to SUSY-invariant true vacua, and show that
there are  severer
constraints on the model parameters.

\section{Transition to the true vacuum at zero temperature}

Let us estimate the quantum transition rate of the
SUSY-breaking to SUSY-invariant vacua in our model by using the
semiclassical approximation \cite{Coleman}. Our model has several
complex scalar fields. We first find a bounce solution by a numerical
method in Ref.~\cite{Konstandin:2006nd} and then estimate the
four-dimensional Euclidean action $S_4$. For the technical details of
our analysis, see Ref.~\cite{Hisano:2007gb}.

We derive a bounce solution whose path is from the SUSY-breaking false
vacuum ($S\sim 0$ and $\Phi_{d}{\bar \Phi}_{d}=\Phi_{l}{\bar
  \Phi}_{l}=0$) to the SUSY-invariant true vacuum ($S={M}/{\kappa}$,
$\Phi_{d}{\bar \Phi}_{d}=0$ and $\Phi_{l}{\bar
  \Phi}_{l}={\mu^2}/{\kappa}$).  We take $S$ and $\Phi_{l}={\bar
  \Phi}_{l}$ real for simplicity, and other fields are vanishing.  The
minimalization of the
$D$-term potential for messengers leads to this configuration
$|\Phi_{l}|=|{\bar \Phi}_{l}|$, and the total potential for the direction
along $\Phi_{l}={\bar \Phi}_{l}$ becomes unstable when $S$ is
increased. Defining $\phi_1\equiv \Phi_{l}={\bar \Phi}_{l}$ and
$\phi_2\equiv S$, we have the following $O(4)$ symmetric Euclidean
action,
\begin{eqnarray} 
\label{eq:action}
 S_4[\phi_i(r)]
 = 
  2\pi^2 \int_0^\infty dr~r^3 
  \left[ \sum_{i=1}^2 \frac{k_i}{2} \left(\frac{d \phi_i}{d r}\right)^2
  + (V_{T=0}(\phi_i)-V_{T=0}(\phi_i^f)) \right] ,
\end{eqnarray}
where $k_1=4$ and $k_2=2$, and $V_{T=0}(\phi_i)$ and
$V_{T=0}(\phi_i^f)$ are the one-loop corrected effective scalar
potential and the false vacuum energy at zero temperature,
respectively. As explained in the previous section, the loop
corrections of mesons $M_a^{\prime}$ and singlets $S_a^{\prime}$
stabilize the SUSY breaking vacuum at $S=0$, while those of the
messengers destabilize it. Thus, the one-loop corrections to the
scalar potential is very important and it should be included in our
analysis.

The SUSY-invariant vacuum has a  flat direction when
$M_d/\kappa_d=M_l/\kappa_l$ in Eq.~(\ref{messenger}). This may enhance
the prefactor $A$ in the transition rate per unit volume, $\Gamma/V=A
\exp(-S_4)$. This situation is very similar to the tunnelings in systems with spontaneous
symmetry breaking discussed in \cite{Kusenko:1995bw}. However, the possible
enhancement would not be significant in the evaluation of the
lowerbound on $S_4$. Here, we approximate the prefactor as $A \sim
\mu^4$, and $S_4\gsim 400$ is required to make the lifetime of the
false vacuum much longer than the age of the universe.

\begin{figure}[t]
\begin{center}
\begin{tabular}{cc}
\includegraphics[scale=0.6, angle = 0]{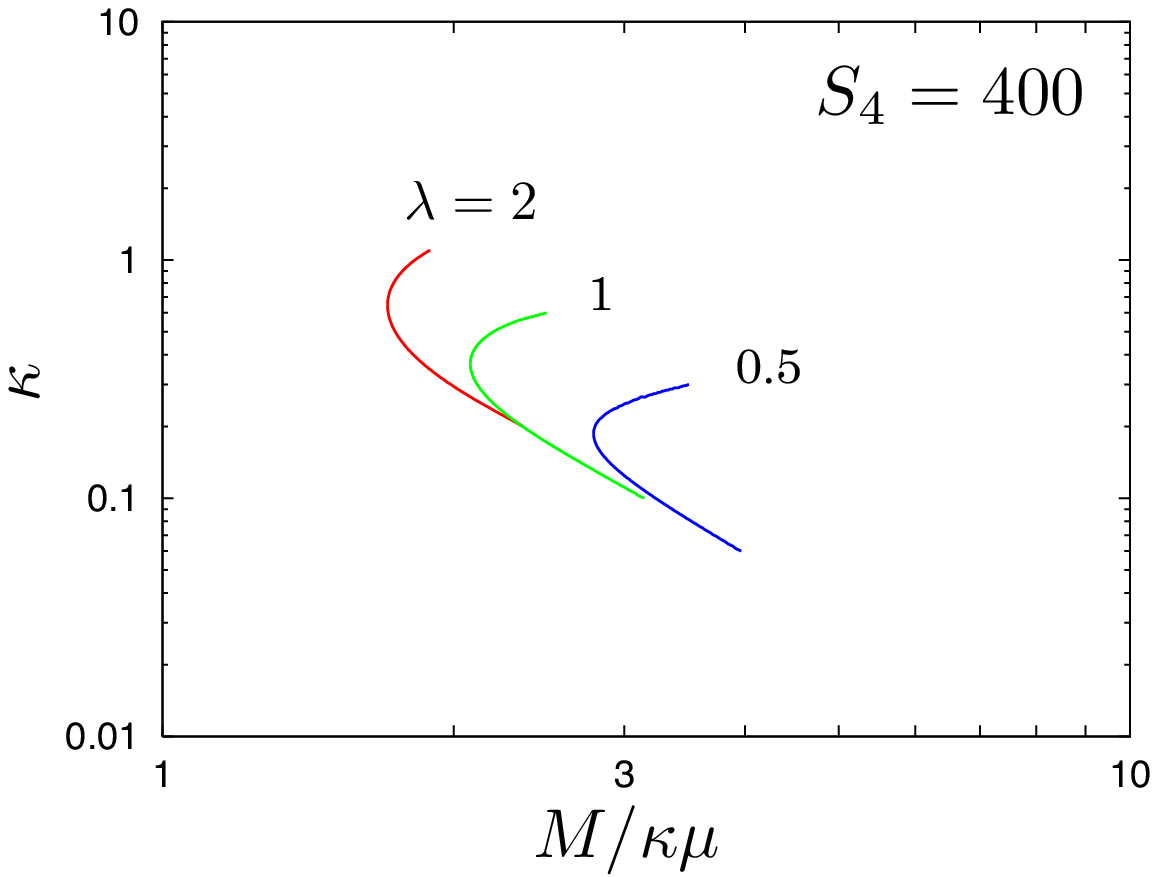}&
\includegraphics[scale=0.6, angle = 0]{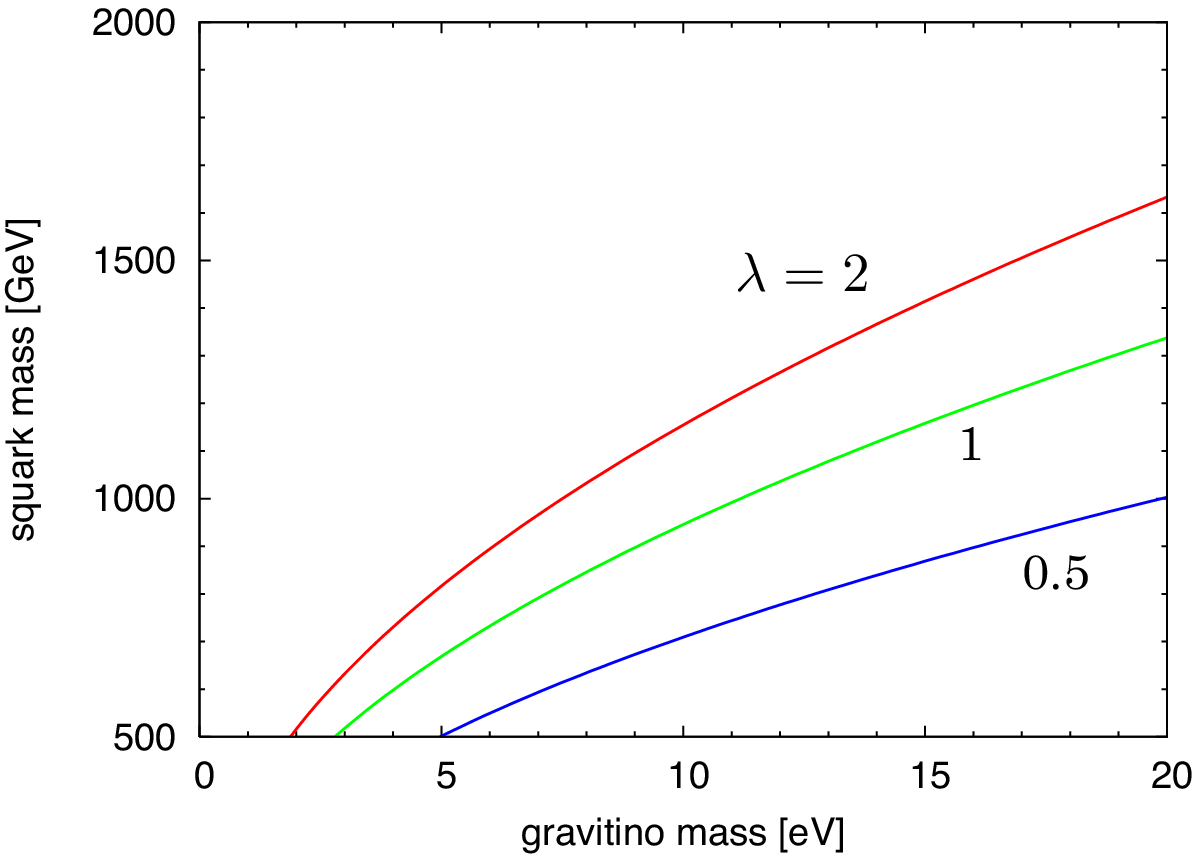}
\end{tabular}
\caption{\label{fig:s4_5} 
(Left) Contour plots for $S_4=400$ on a plane of $M/\kappa\mu$
$(\mu^2\equiv\lambda \Lambda^2)$ and $\kappa$ for $\lambda=0.5,~1,~2$.
On the right-handed sides of the lines $S_4$ is larger than 400.
(Right) Squark mass upperbounds from quantum stability of the SUSY-breaking 
vacuum as functions of gravitino mass.
}
\end{center}
\end{figure}

In Fig~\ref{fig:s4_5} (left), we show contour plots for $S_4= 400$ on
a plane of $\kappa$ and $M/\kappa\mu$ for 
$\lambda=$2, 1, 0.5. The lines follow almost the boundary of the classical
stability conditions of the SUSY-breaking vacuum in Eqs.~(\ref{cond1},
\ref{cond2}); $(M/\kappa\mu)^{-2}\lsim \kappa \lsim \lambda^{3/2}(M/\kappa
\mu)$.  Notice that larger $\lambda$ increases the  mass squared $m_S^2$ 
and hence it makes smaller $M/\kappa\mu$ to be allowed.

In the parameter regions away from the lines toward the right, the
parameter dependence of $S_4$ is understood from an approximate
estimate based on Ref.~\cite{Duncan:1992ai}. In the approximation the
vacuum transition rate is calculated for a triangle-shaped potential
of one real scalar field. In our model, the true vacuum becomes more
far from the false one for larger $M / \kappa \mu$, while the
potential barrier is not high compared with the energy gap between
true and false vacua.  In this case, $S_4$ is approximately given as
\begin{eqnarray}
 S_4 \simeq 8\pi^2 \left(\frac{M}{\kappa \mu}\right)^4 \ .
\end{eqnarray}
Then, when $M/\kappa \mu$ is larger, the SUSY-breaking vacuum is more
stable against the quantum vacuum transition.

In Fig~\ref{fig:s4_5} (right), the upperbounds on the squark masses 
are shown for several values of $\lambda$ as
functions of the gravitino mass. It is found from Eq.~(\ref{eq:gravitino}) that smaller
$M/\kappa \mu$ allows larger squark masses with $m_{3/2}$ fixed, while
$M/\kappa \mu$ is bounded from below due to the quantum stability of
the SUSY-breaking false vacuum. Thus, it is found that the squark masses
have to be smaller than 1.5 (1.2,~0.9)~TeV for $\lambda$=2 (1,~0.5)
for $m_{3/2}<16$~eV.

\section{Thermal  transition in the early universe}

Now, we discuss about thermal transitions of the SUSY-breaking to
SUSY-invariant vacua in the early universe. At very high temperatures,
the SUSY-breaking vacuum is automatically selected due to the thermal
correction to the scalar potential in our model as shown
below. However, it depends on various parameters in the model whether
the false vacuum survives or not until temperature falls much below
$\sim \mu$.

Let us explain the thermal history of the universe in our model. When
the temperature of the universe $T$ falls below $\sim 4\pi \Lambda$, the
hidden $SU(2)$ gauge interaction in the SUSY-breaking sector becomes strong,
and the $SU(2)$ quarks are confined so that only the mesons $M_a^\prime$
become the dynamical degree of freedom in the SUSY breaking sectors
together with singlets $S_a^\prime$ and $S$.

At $T \gsim \mu$, the scalar potential of $S$
and the messenger leptons, $\Phi_l$ and $\bar \Phi_l$, is
approximately given by
\begin{eqnarray}
V_T&\simeq& V_{T=0}+
\frac18 (3g_2^2+g_Y^2+2 \kappa^2)T^2 (|\Phi_l|^2+|\bar\Phi_l|^2)
\nonumber\\&&
+\frac58 \lambda^2T^2|S|^2 + \frac54 T^2|\kappa S-M|^2 \ .
\label{tempra}
\end{eqnarray}
Here, $g_2$ and $g_Y$ are $SU(2)_L$ and $U(1)_Y$ gauge coupling
constants, respectively.  The thermal corrections by the SSM gauge
interactions lift the scalar potential of the messengers.  The
messenger quarks, $\Phi_d$ and $\bar \Phi_d$, have larger thermal mass
terms than the messenger leptons due to
the $SU(3)_C$ interaction, though they are omitted in
Eq.~(\ref{tempra}).  The field values of the messengers are
zero at the minimum of the scalar potential. The SUSY-invariant vacuum is hidden
due to the thermal correction.

In addition, the interaction of mesons $M_a^{\prime}$ with
$S$ makes the minimum around $S=0$ in the thermal
potential.  Since the messenger masses are given as $M-\kappa S$, the
messengers would make a local minimum around $S\simeq
M/\kappa$. However, the local minimum does not appear as far as
$\lambda \gsim \kappa$, and the global minimum of the potential at
$S\simeq 0$ and $\Phi_{d/l}=\bar{\Phi}_{d/l}=0$ is close to the
SUSY-breaking vacuum at zero temperature\footnote{
Exactly speaking, when $M/\kappa \gg \lambda \mu$, the approximation in
Eq.~(\ref{tempra}) is not valid and a local minimum around $S\simeq
  M/\kappa$ appears at $T\gsim \mu$. The thermal potential
  by mesons and singlets cannot lift the potential of $S$ much enough
  to hide the local minimum. If the subcritical bubbles of the local
  minimum are created by thermal hopping \cite{Gleiser:1991rf} and
  they survives at $T\lsim \mu$, the phase transition
  to the true vacuum would be efficient.  In this paper we are interested in
  smaller $M/\kappa\mu$, which leads to the light gravitino. Then, we
  do not discuss this case furthermore.  }.

When the temperature falls below $\sim \mu$, the stability of the
SUSY-breaking vacuum becomes weaker.  The SUSY true vacuum starts to
appear, since the thermal potential by the SSM gauge interactions
cannot hide it. It is found from Eq.~(\ref{tempra}) that the critical
temperature $T_c$, at which the SUSY-breaking and SUSY-invariant vacua
are degenerate, is $\lsim \sqrt{8\kappa/3 g_2^2}\times \mu$.  In addition,
mesons $M_a^\prime$ and the singlets $S_a^\prime$ are gradually
decoupled from the thermal bath so that the interaction of the
messengers become relatively stronger.  Notice here that the masses of
the mesons and the singlets are $\sqrt{\lambda}\mu$.

After the mesons are decoupled, the first term in the second line of
Eq.~(\ref{tempra}) disappears so that $S$ moves from $S\simeq 0$
toward $M/\kappa$ $(S\simeq (4\pi^2(T/\mu)^2 \kappa^2/\lambda^3)\times
(M/\kappa))$.  If $S$ approaches close to $M/\kappa$, $S$ and
messengers would roll into the SUSY-invariant vacuum. Thus, $\kappa$ should be
suppressed so that the universe is trapped in the SUSY-breaking
vacuum.

\begin{figure}[t]
\begin{center}
\begin{tabular}{cc}
\includegraphics[scale=0.6, angle = 0]{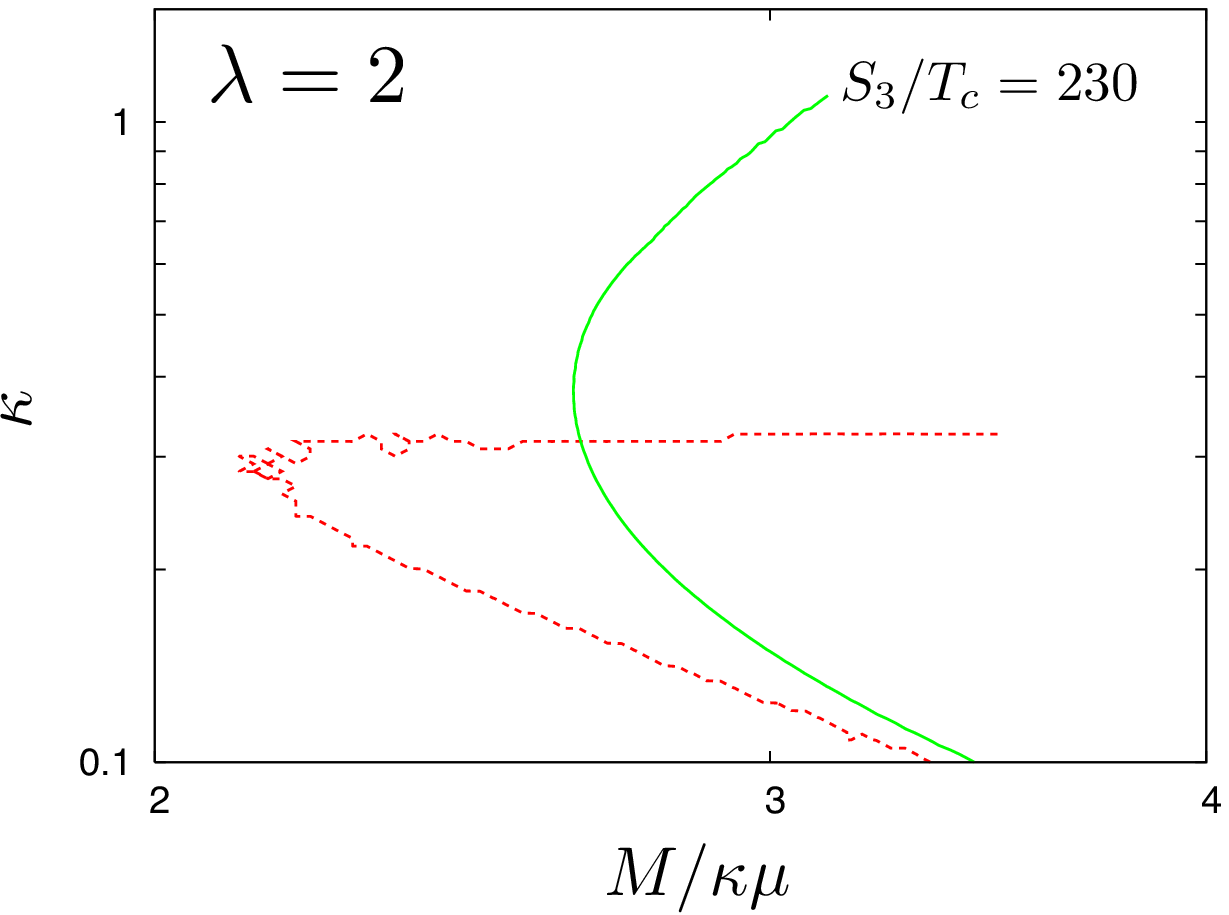}&
\includegraphics[scale=0.6, angle = 0]{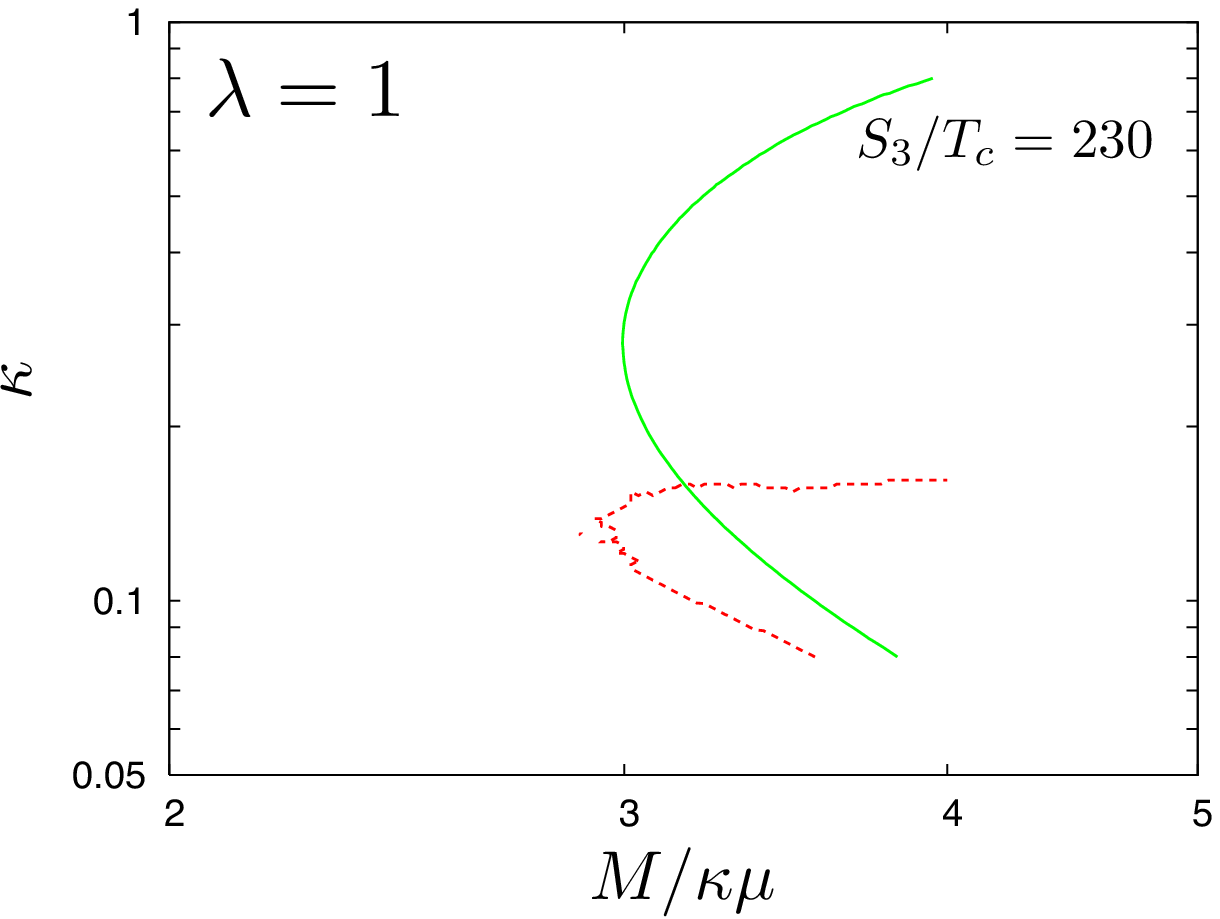}\\
\includegraphics[scale=0.6, angle = 0]{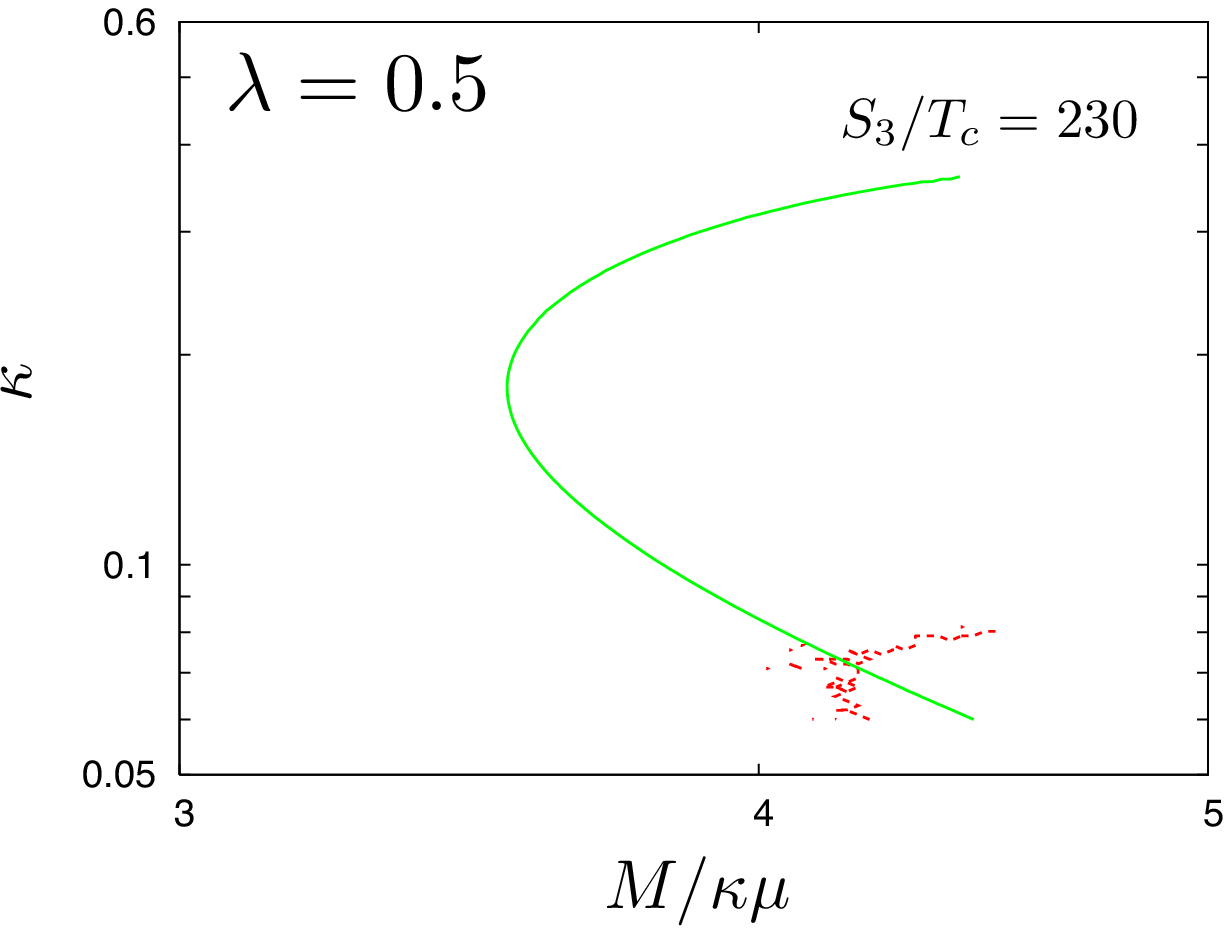}&
\end{tabular}
\caption{\label{fig:s3_5} 
Solid (green) lines are for $S_3/T_c=230$ in plans
of $\kappa$ and $M/\kappa\mu$ for $\lambda=2,1,0.5$. 
In regions except for the right-handed sides of dashed (red) lines, the 
SUSY-breaking vacuum becomes unstable in lower
temperature than $\sim \mu$, at which the universe rolls down to the SUSY-invariant vacuum. 
}
\end{center}
\end{figure}

\begin{figure}[t]
\begin{center}
\begin{tabular}{c}
\includegraphics[scale=0.6, angle = 0]{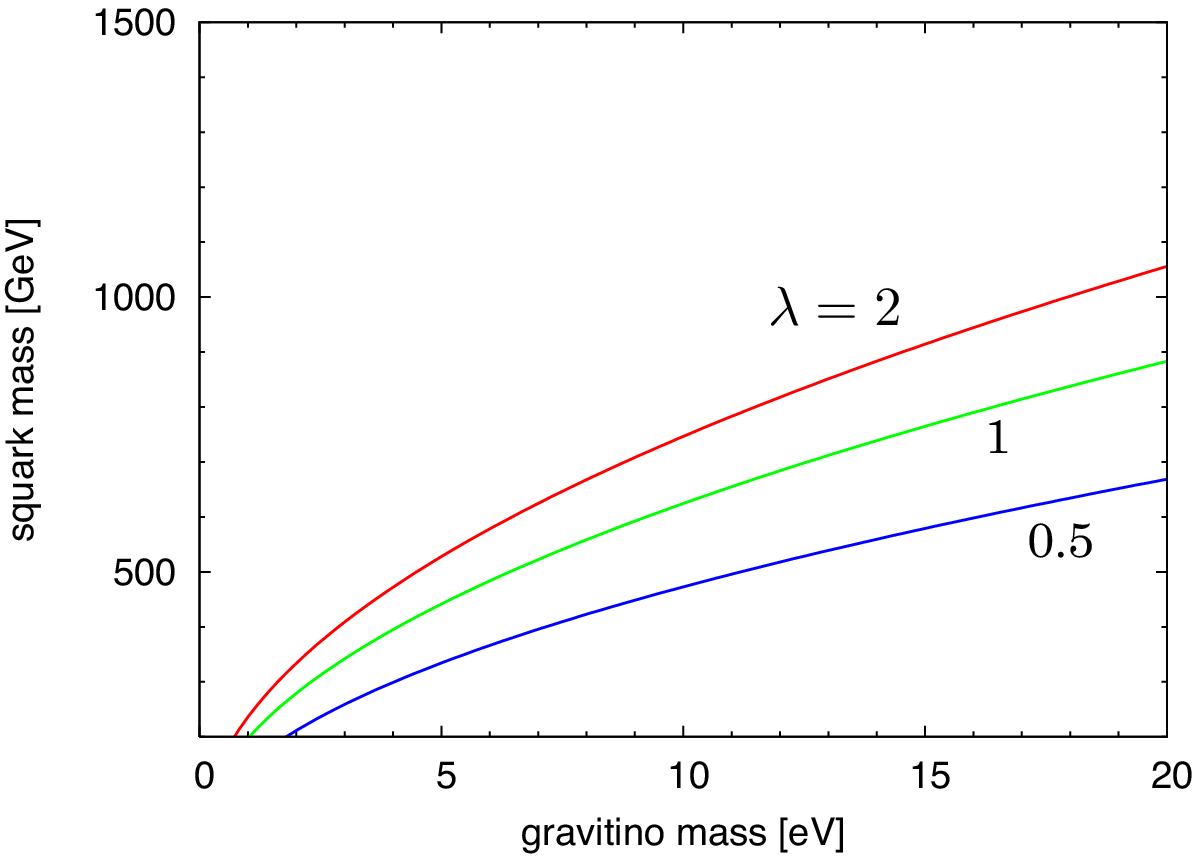}
\end{tabular}
\caption{\label{fig:3} 
Squark mass upperbounds as functions of the gravitino mass for $\lambda=0.5,
~1,~2$ from the stability of the SUSY-breaking vacuum in the thermal bath of 
the early universe.
}
\end{center}
\end{figure}

Now we estimate the thermal transition rate from the SUSY-breaking to
SUSY-invariant vacua. In the case of a finite temperature, the transition rate
is determined by the three-dimensional Euclidean action $S_3$ as
\cite{Linde:1981zj}
\begin{eqnarray}
\Gamma_3&\simeq& T^4\left(\frac{S_3}{2\pi T}\right)^{3/2} e^{-\frac{S_3}{T}} \ .
\end{eqnarray}
Most of the present universe remains in the false (SUSY-breaking)
vacuum when ${S_3}/{T}\gsim 230$ \cite{Linde:1981zj,Guth:1981uk}.  

When evaluating $S_3$, we choose a bounce solution whose path is from
the SUSY-breaking vacuum ($S\sim 0$ and $\Phi_{d}{\bar
  \Phi}_{d}=\Phi_{l}{\bar \Phi}_{l}=0$) to the SUSY-invariant vacuum
($S={M}/{\kappa}$, $\Phi_{d}{\bar \Phi}_{d}=0$ and $\Phi_{l}{\bar
  \Phi}_{l}={\mu^2}/{\kappa}$). The SUSY-invariant vacuum with
$\Phi_{d}{\bar \Phi}_{d}\ne 0$ suffers from the stronger thermal
correction due to the $SU(3)_C$ interaction as mentioned above, and it
is more hidden at the low temperatures than the vacuum with $\Phi_{d}{\bar
  \Phi}_{d}= 0$ and $\Phi_{l}{\bar \Phi}_{l}={\mu^2}/{\kappa}$. This
is a reason why we consider the above vacuum transition.

The $O(3)$ symmetric Euclidean action is given by
\begin{eqnarray} 
\label{eq:action3}
 S_3[\phi_i(r)]
 = 
  4\pi \int_0^\infty dr~r^2 
  \left[ \sum_{i=1}^2 \frac{k_i}{2} \left(\frac{d \phi_i}{d r}\right)^2
  + (V_{T}(\phi_i)-V_{T}(\phi_i^f)) \right] \ .
\end{eqnarray}
Here, $V_{T}(\phi_i^f)$ is the false vacuum energy at
$T$.  The scalar potential is almost flat around the critical
temperature $T_c$. Then, we assume in evaluating $S_3$ that
$V_T(\phi_i)$ is dominated by the zero-temperature potential, while we
derive $T_c$ from the potential including thermal corrections. We use
the formula for the thermal potential given in
Ref.~\cite{Dolan:1973qd}. We estimate $S_3$ by the numerical method in
Ref.~\cite{Konstandin:2006nd}, again.

In Fig.~\ref{fig:s3_5} we show that contour plots for $S_3/T_c=230$ on
planes of $\kappa$ and $M/\kappa\mu$ for $\lambda=$2, 1, 0.5 by solid (green)
lines.  The left-handed sides of the lines, in which $S_3/T_c<230$, are
disfavored. It is found from the triangle approximation that $S_3$ is
approximately given by $8\pi/3(M/\kappa\mu)^3\mu$ in the regions away
from the lines to the right-handed sides.

In addition, in regions except for the right-handed sides of dashed (red)
lines, the SUSY-breaking false vacuum becomes destabilized in lower
temperature than $\sim \mu$ so that the universe rolls down to the
SUSY-invariant vacuum. When $\kappa$ is larger, the messengers may destabilize
the false vacuum as mentioned above. This excludes broad regions in
the parameter space.

The stability of the SUSY-breaking vacuum in the thermal bath of the
early universe gives a stronger constraint on $M/\kappa\mu$. We
translate it to the squark mass upperbounds as functions of the
gravitino mass for $\lambda=0.5,~1,~2$ in Fig.~\ref{fig:3}. When
$m_{3/2}<16$~eV, the squark masses have to be smaller than $
1~(0.8,~0.6)$~TeV for $\lambda=2~(1,~0.5)$. It is found after
taking the mass functions $f$ and $g$ and the renormalization-group
effects into consideration that the
squark mass $m_{\tilde{q}}<1$~TeV corresponds to the gluino mass
$m_{\tilde{g}} \lsim 1~{\rm TeV}$ in our model.

Notice that we imposed $S_3/T>230$ at $T=T_c$ while using the
zero-temperature scalar potential in the evaluation of $S_3$. If we
evaluate $S_3$ as a function of $T$ using the potential including
thermal correction and impose $S_3/T>230$ for $T\le T_c$, the
constraints on the squark masses would be weakened. We consider from studies
of some sample points in our model that the correction on the squark
mass upperbounds is at most $O(10)\%$.

\section{Conclusions and discussion}

\begin{figure}[t]
\begin{center}
\begin{tabular}{c}
\includegraphics[scale=0.6, angle = 0]{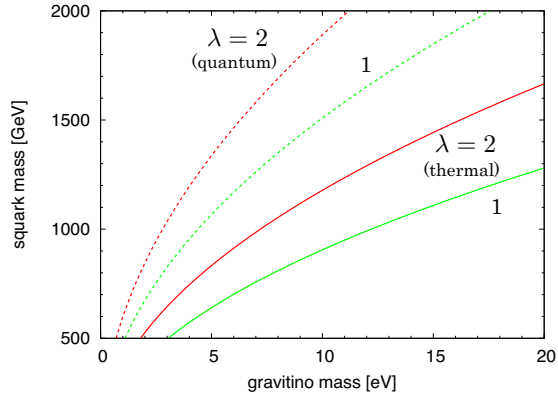}
\end{tabular}
\caption{\label{fig:squark10} 
Squark mass upperbounds as functions of gravitino mass for $\lambda=1,~2$.
Here, we assume the messenger model with ${\bf 10}+{\bf 10^\star}$.
Dashed (solid) lines come from stability of the SUSY-breaking 
vacuum against quantum (thermal) transition. 
}
\end{center}
\end{figure}

In this paper we have discussed the vacuum transitions at zero
temperature and at high temperatures in the early universe by assuming
the IYIT model for the gauge mediation. This model accommodates a
gauge mediation with the light gravitino of mass $m_{3/2} \lsim
16$~eV, but the SUSY particle masses in the SSM model are bounded from
above to maintain the thermal and quantum stabilities of the SUSY-breaking
false vacuum. We have, in fact, found that $m_{\tilde{q}} \lsim 1~{\rm TeV}$ and
1.5~TeV for the thermal and quantum stabilities, respectively, when
imposing $m_{3/2} \lsim 16$~eV.

In the rest, we discuss possibilities to weaken the bounds.  The
bounds we have derived in this paper are based on the minimal messenger model,
in which $ {\bf 5}+{\bf 5^\star}$ in the $SU(5)_{\rm GUT}$ are
introduced. An alternative choice of the messengers is a pair of
messengers belonging to ${\bf 10}+{\bf 10^\star}$. In
Fig.~\ref{fig:squark10}, the upperbounds for the squark masses are
shown as functions of the gravitino mass in the messenger model with
${\bf 10}+{\bf 10^\star}$. The solid (dashed) lines come from the
thermal (quantum) stability for $\lambda=1,~2$. While increase of the
number of the messengers makes stability of the SUSY-breaking vacuum
weaker, the gaugino and sfermion masses are enhanced by $3$ and
$\sqrt{3}$, respectively. It is found that the the upperbound on
squark masses from the thermal stability is increased to 1.5 TeV. This
corresponds to the upperbound on the gluino mass $m_{\tilde{g}} \lsim
2.0$ TeV.

For the SUSY-breaking sector, we can extend the hidden gauge symmetry to
$Sp(2N)$ gauge group. The number of mesons and singlets, $M_a^\prime$
and $S_a^\prime$, is $2N^2+3N$, while the Yukawa coupling of mesons
with $S$ is suppressed by $1/\sqrt{N}$. Then, $m_S^2$ is enhanced by
$N^{1/2}$, and the larger $N$ stabilizes the SUSY-breaking vacuum
more strongly.  In addition, the mass of mesons and singlets is scaled
by $N^{-1/4}$.  When $N\gg 1$, the thermal correction to the scalar
potential by the light mesons is more efficient at low temperatures than in the
$SU(2)$ case.  Thus, large $N$ model allows larger squark masses than the
original $SU(2)$ model.

Finally, we should stress that an improvement of the gravitino mass
bound \cite{Viel:2005qj} by a factor of 2 may give a serious problem
for the present gauge mediation model.

\section*{Acknowledgement} This work was supported by World Premier
International Center Initiative (WPI Program), MEXT, Japan.  The works
of JH and TTY was also  supported in part by the Grant-in-Aid for Science
Research, Japan Society for the Promotion of Science
(No.~20244037 and No.~2054252 for JH and No.~1940270 for TTY).

\end{document}